\documentclass[11pt]{article}
\newcommand{\Comment}[1]{{}}
\usepackage{amsfonts,amsmath, amsmath,amssymb,slashed}
\usepackage[textwidth = 430 pt, textheight = 630 pt]{geometry}

\Comment{\usepackage{color}
\definecolor{MyDarkBlue}{rgb}{0.15,0.15,0.45}
\usepackage[linktocpage=true]{hyperref}
\hypersetup{
colorlinks=true,
citecolor=MyDarkBlue,
linkcolor=MyDarkBlue,=
urlcolor=MyDarkBlue,
pdfauthor={},
pdftitle={},
pdfsubject={hep-th}
}

\usepackage[numbers,sort&compress]{natbib}
\usepackage{hypernat}}
\usepackage{graphicx}
\usepackage{cite,color}

\def\e{{\rm e}}
\def\bz{\bar{z}}
\def\ep{\epsilon}

\def\g{\gamma}

\def\D{\mathcal{D}}
\def\tr{{\rm Tr}}

\def\d{\partial}
\def\U{\mathcal{U}}

\newcommand{\be}{\begin{equation}}
\newcommand{\bea}{\begin{eqnarray}}

\newcommand{\ee}{\end{equation}}
\newcommand{\eea}{\end{eqnarray}}

\newcommand{\nn}{\nonumber}

\newcommand{\half}{\frac{1}{2}}

\begin{document}

\vspace{10pt}


\begin{center}
{\LARGE \bf{\sc Asymptotic Symmetries of Yang-Mills with Theta Term and Monopoles}}
\end{center} 
 \vspace{1truecm}
\thispagestyle{empty} \centerline{
{\large \bf {\sc Carlos Cardona}}\footnote{E-mail address: \Comment{\href{mailto:cargicar@gmail.com}}{\tt cargicar@gmail.com}}
 }
\vspace{.5cm}

\centerline{{\it Physics Department, Universidad del Valle\footnote{Temporary affiliation }}}\centerline{{\it Calle 13 Number 100-00, Santiago de Cali, Colombia.}}
\vspace{0.5cm}

\vspace{1truecm}

\thispagestyle{empty}

\centerline{\sc Abstract}

\vspace{.4truecm}

\begin{center}
\begin{minipage}[c]{380pt}
{In this short note we suggest that the singular behavior of large gauge transformations preserving the vacuum at null infinity in Yang-Mills theory implies monopoles into the bulk, as well as that the inclusion of a theta term induces a decoupling between holomorphic and anti-holomorphic currents associated to those large gauge transformations  }
\end{minipage}
\end{center}

\vspace{.5cm}

\setcounter{page}{0}
\setcounter{tocdepth}{2}



\setcounter{tocdepth}{2}
\tableofcontents

\newpage
\setcounter{equation}{0}

\section{Introduction}

Recently there has been a renewed interest in the study of the asymptotic symmetries that preserve the Minkowskian structure of asymptotically flat space-times. It has been known for a long time \cite{Bondi:1962,Sachs:1962} that this is a infinite symmetry,  composed by Lorentz symmetry plus the so-called supertranslations. In retarded or advanced coordinates the asymptotic region has the topology of  $\mathbb{R}\times S^2$, where supertraslations act on the null direction $\mathbb{R}$ and are infinite, and the Lorentz group corresponds to the $SL(2,\mathbb{C})$ preserving the structure of $S^2$. This set of symmetries is known as Bondi-van der Burg-Metzner-Sachs (BMS). However, very recently it has been shown that the BMS group can be enhanced by allowing singular transformations on the sphere at spatial infinity \cite{Barnich:2011ct,Barnich:2011mi}. Using this fact Strominger and Cachazo have found a new Weinberg's soft theorem that contains an additional subleading soft factor that depends on the angular momentum of the particle emitting the soft graviton \cite{Cachazo:2014fwa}. Subleading soft factors for double soft emission in a variety of theories have also been found very recently in \cite{Cachazo:2015ksa}.

By considering the analogous situation in Yang-Mills theory, Strominger and collaborators have managed to get the soft theorem for gluons and photons \cite{Strominger:2013lka, He:2015zea, He:2014cra} as the Ward identity corresponding to the asymptotic singular gauge symmetry, which are the analogous of the asymptotic  BMS diffeomorphisms in gravity. The subleading factors correcting the Weinberg's soft theorem in Yang-Mills were obtained in \cite{Casali:2014xpa}.

As asymptotic gauge transformations act on the two dimensional $S^2$, it is natural to expect that the asymptotic symmetry $SL(2,\mathbb{C})$ gets enhanced to a Kac-Moody symmetry as long as singular transformations are allowed, which was actually shown in \cite{He:2015zea} for Yang-Mills. However, it was found that the usual holomorphic and anti-holomorphic currents do not decouple, and hence is only possible to define a single holomorphic (or anti-holomorphic) current algebra.

Asymptotic symmetries at the boundary are particularly important in the context of the AdS/CFT correspondence, whose roots can be traced back to the work of Brown and Henneaux where it was observed that for a theory of gravity in three dimensional $AdS$, the algebra of asymptotic symmetries at the boundary is Virasoro  \cite{Brown:1986nw}. This has also played a central role in the understanding of the $AdS_3/CFT_2$ correspondence.

In this note, we would like to suggest that by allowing singular gauge transformations at infinity in Yang-Mills theory, one should induce monopoles. Additionally, by adding a theta term to the non-abelian Yang-Mills Lagrangian, we will show that the surface term coming from it allows us to define two independent holomorphic and anti-holomorphic current algebras at different values of the theta parameter, just as in two-dimensional WZW models, which was actually the expected situation.

This paper is organized as follows, in section two we discuss how the singular gauge transformations at null infinity imply the existence of monopoles at the bulk for a $U(1)$ theory, then in section three we show the same phenomena for the non-abelian case, but we also discuss how the inclusion of a theta term induces a decoupling between the holomorphic and anti-holomorphic currents associated to the large gauge transformations.
\section{Large Singular Gauge Transformations and Monopoles in Maxwell Theory}
Let us star considering a $U(1)$ gauge theory coupled to some massless charged scalars,
\be
\mathcal{S}=-\frac{1}{4e^2}\int d^4x\sqrt{g}\left(F_{\mu\nu}F^{\mu\nu}+\D_{\mu}\Phi\D^{\mu}\Phi\right)
\ee
We rewrite the space-time metric $ds^2=\eta_{\mu\nu}d x^{\mu}d x^{\nu}$ in terms of retarded coordinates
\be 
v=t-r,\quad x^1+ix^2=\sqrt{2\g_{z\bz}}rz,\quad x^3=
\sqrt{\frac{\g_{z\bz}}{2}}(1-z\bz)\,.
\ee
In the particular case of the stereographic projection, we can take on the north chart $z=\e^{i\phi}\tan(\theta/2)$, leading to the round metric on the $S^2$ sphere $\g_{z\bz}=\frac{2}{(1+z\bar{z})^2}$. The space-time metric looks like,
\be
ds^2=-dv^2-2dvdr+2r^2\g_{z\bz}dzd\bz\,. 
\ee
The spatial infinity $r=\infty$ is commonly denoted $\mathcal{I}^+$ and has the topology $S^2\times\mathbb{R}$  with coordinates $(v,z,\bz)$ and has its null boundaries at $v=\pm\infty$ which are denoted by $\mathcal{I}^+_{\pm}$.
Similarly we have the advanced coordinates
\be 
u=t+r,\quad x^1+ix^2=-\sqrt{2\g_{z\bz}}rz,\quad x^3=
-\sqrt{\frac{\g_{z\bz}}{2}}(1-z\bz)\,.
\ee 
which give rise to
\be
ds^2=-du^2-2dudr+2r^2\g_{z\bz}dzd\bz\,. 
\ee
Its topology is again $S^2\times\mathbb{R}$ and the null boundary $v=\pm\infty$ is denoted by $\mathcal{I}^-_{\pm}$.

We would like to consider configurations approaching the vacuum at the boundaries ${\cal I}^{\pm}_{\pm}$,
\be\label{vacuum}
F_{\mu\nu}=0,\quad \D_{\mu}\Phi=0\,. 
\ee
We work in the radial and asymptotic retarded gauge 
\be\label{gauge}
A_r=0,\quad A_v|_{{\cal I}^+}=0\,,\ee
and expand the fields in the asymptotic region as in \cite{Strominger:2013lka},
\bea\label{expansion}
A_z(r,v,z,\bz)&=&\sum_0^{\infty}\frac{A_z^{(n)}(v,z,\bz)}{r^n}\,,\\
j^{E}_v(r,v,z,\bz)&=&\sum_0^{\infty}\frac{j_v^{(n)}(v,z,\bz)}{r^{n+2}}\,. 
\eea
Using the Maxwell equations in the gauge (\ref{gauge})\footnote{As in  \cite{He:2014cra,Strominger:2013lka} we take $j^{E}_{\nu}=(j^{E}_{u},0,0,0)$ for the sake of simplicity.} $\nabla^{\mu}F_{\mu\nu}=j^{E}_{\nu}$, we get \cite{Strominger:2013lka, He:2014cra}
\bea
\d_v(\d_z A_{\bz}+\d_{\bz}A_z)&=&er^2\gamma_{z\bz} j^{E}_v\,,\label{eom}\\
\d_r(\d_z A_{\bz}+\d_{\bz}A_z)&=&0\,,\nn\\
r^2\d_r(\d_r A_z-2\d_v A_z)])&=&0\nn
\eea
Also, the gauge conditions (\ref{gauge}) left a residual gauge freedom at ${\cal I}^+$ generated by an arbitrary function $\ep(z,\bz)$ and hence the vacuum conditions (\ref{vacuum})  implies a flat field strength $F_{z\bz}=0$ at spatial infinity,
\be A_z|_{{\cal I}^+_{\pm}}=A_z(\infty,z,\bz)=\partial\ep(z,\bz)\,.\ee
Putting this condition into the equations of motion (\ref{eom}) and using the expansion (\ref{expansion}) one can write the leading field $A_z^{(0)}$  and $A_{\bz}^{(0)}$ in terms of the leading current as,
\bea\label{abz}
\d_{\bz} A^{(0)}_z&=&\frac{e}{2}\gamma_{z\bz}\int dv j^{(0)}_v(v,z,\bz)\,.\\
\d_{z} A^{(0)}_{\bz}&=&0\,.
\eea
For a set of massless point electric particles crossing ${\cal I}^+$ we use the current,
\be\label{pointcurrent}
j^{(0)}_v=\sum_{k}q_k\delta(v-v_k)\frac{\delta(z-z_k)}{\gamma_{z\bz}}\,,\ee 
such that (\ref{abz}) is solved by\footnote{At the sphere this solution is unique, unlike to surfaces of higher genus. I thank to Humberto Gomez for point it out to me.},

\be\label{singpote}
A^{(0)}_z(\infty,z,\bz)=\frac{e}{2}\sum_{k}\frac{q_k}{z-z_k}\,.
\ee
Since it is a singular gauge potential, the flux over the asymptotic sphere of the magnetic field is non-zero, and hence there should be Dirac monopoles. The usual Dirac monopole discussed in the literature has the following form in stereographic coordinates\footnote{ This is the form over the north chart},
\be
A_{+}=\frac{1}{4\pi r}\frac{\e^{i\phi}}{z}\,.
\ee
the potential is in this case singular at the south pole of $S^2$ which in turn induce a magnetic flux over the sphere. The potential (\ref{singpote}) is then the usual Dirac monopole at conformal infinity and with $\phi=0$.
 Let us compute the magnetic flux  produced by the potential (\ref{singpote}) over the $S^2$ at null infinity,

\be \int_{S^2}B=\int_{\d S^2} A_z=\frac{e}{2}\sum_{k}q_k\oint_{{\cal C}_k}\frac{dz}{z-z_k}=\frac{e}{2}\sum_{k}q_k\,.\ee
Since the magnetic flux should be an integer we obtain the Dirac quantization condition\footnote{For a total magnetic charge $g=1$}.
\be Q=4\pi n \,,\ee
where we have defined the total charge as $Q=\sum_keq_k$.
The usual interpretation is that some Dirac strings pinch the sphere at the points $z_k$.






\section{Non-Abelian Gauge Theory with Theta Term}
In this section we re-do the analysis of the previous section for a Yang-Mills theory with gauge group ${\cal G}$ and scalar matter, however our main concern is to investigate the effect of a theta term on the asymptotic symmetries. Let us start reviewing the case without theta term in order to elucidate the effect of the topological contribution. Although we believe the results are more general, let us take the following action for the sake of clarity,
\be\label{ggaction}
\mathcal{L}=-\frac{1}{4}\tr(F_{\mu\nu}F^{\mu\nu})-\half\D_{\mu}\phi\D^{\mu}\phi-V(\phi)\,.
\ee
Our conventions for this theory are given by
$\tr[T^aT^b]=+\delta^{ab}$, $\D^{\mu}=\d^{\mu}-ig_{YM}[A^{\mu},.]$ and $[D_{\mu},D_{\nu}]=-ig_{YM}F_{\mu\nu}$ with
\be F_{\mu\nu}=\D_{[\mu}A_{\nu]}-ig_{YM}[A_{\mu},A_{\nu}]=F^a_{\mu\nu}T^a\,.\ee
We would like to consider a configuration such that the fields approach the vacuum at $\mathcal{I}^{\pm}_{\pm}$, where the fields will take the following values,
\be\label{vacum}
{F_{\mu\nu}^a=0},\quad \D\phi^a=0,\quad V(\phi)=0\,. 
\ee
In the particular Georgi-Glashow model where ${\cal G}=SU(2)$ and $V=\frac{g_{YM}^2}{4}(\phi^2-\phi_0^2)^2$, the vacuum breaks the gauge symmetry down to $U(1)$, and the function $\phi(z,\bz)$ maps the $S^2$ at null infinity $(r=\infty,u=\infty)$ to the $S^2$ which leaves the vacuum $\phi^2=\phi_0^2$  invariant. Mappings from $S^2$ to $S^2$ are characterized by the integers $\Pi_2(S^2)=\mathbb{Z}$, moreover, the symmetry breaking at null infinity implies the existence of 't Hooft-Polyakov monopoles \cite{'tHooft:1974qc,Polyakov:1974ek} the same way the singular asymptotic behavior in the abelian case implies Dirac monopoles, whose magnetic charges are quantized corresponding to the integers $\Pi_2(S^2)$.

As in the abelian case we would like to work in the asymptotic retarded gauge 
\be\label{retarded} A_v|_{\mathcal{I}^+}=A_r=0\,. 
\ee
Such that the leading terms of the field strength at $\mathcal{I}^+$ are
\be F_{vr}=0,\quad F_{vz}=\d_uA_z,\quad F_{z\bz}=\d_zA_{\bz}-\d_{\bz}A_z-ig_{YM}[A_z,A_{\bz}]\,.\ee
This implies that at $\mathcal{I}^+$ there is still a residual gauge freedom given by transformations which are independent of $(v,r)$, i.e, 
\be A_{\mu}|_{\mathcal{I}^+}\to A_{\mu}|_{\mathcal{I}^+}+i\U(z,\bz)\d_{\mu}\U(z,\bz)^{-1}\,.\ee
with $\U(z,\bz)\in {\cal G}$. Those transformations have been called large gauge symmetries in \cite{He:2015zea}.

The equations of motion of the action (\ref{ggaction}) are
\bea
\D^{\mu}F_{\mu\nu}&=&g_{YM}\left([\D_{\nu}\phi,\phi]+j^{E}_{\nu}\right)\equiv g_{YM}J^{E\phi}_{\nu}\nn\\ \D_{\mu}\D^{\mu}\phi&=&-V(\phi)\,.
\eea
Under conditions (\ref{vacum}) in the retarded gauge (\ref{retarded}) the above equations takes the form,
\bea
\d_v(\d_z A_{\bz}+\d_{\bz}A_z+ig_{YM}[A_z,A_{\bz}])&=&g_{YM}r^2\gamma_{z\bz} j^{E}_v\,,\label{veom}\\
\d_r(\d_z A_{\bz}+\d_{\bz}A_z+ig_{YM}[A_z,A_{\bz}])&=&g_{YM}r^2 \gamma_{z\bz}  j^{E}_r\,,\\
r^2\d_r(\d_r A_z-2\d_v A_z)+\d_z(\gamma^{z\bz}ig_{YM}[A_z,A_{\bz}])&=&g_{YM}r^2 \gamma_{z\bz}  j^{E}_z
\eea

Once the vacuum $F_{\mu\nu}=0$ is reached at spatial infinity $S^2$, the condition $F_{z\bz}=\d_zA_{\bz}-\d_{\bz}A_z-ig_{YM}[A_z,A_{\bz}]=0$ implies that $A_z$ and $A_{\bz}$ become pure gauge 
\be A_z|_{\mathcal{I}^+_+}=i\U\d_{z}\U^{-1}\,.\ee
Using this fact, one can solve (\ref{veom}) for the  leading modes $ A^{(0)}_{z}$ in terms of the leading color current as in the abelian case, i.e, for a set of massless point particles crossing ${\cal I}^+$ one can write,
\be 
\d_{\bz}A_z^{(0)}(\infty,z,\bz)=\frac{g_{YM}}{2}\sum_{k}\epsilon_k(z,\bz)\delta^2(z-z_k)\,,
\ee
where $\epsilon(z,\bz)$ corresponds to the infinitesimal generator for large gauge transformations in the representation of the charged particle insertion at $z_k$.
The potential is singular at the point insertions which induce a ``magnetic" color flux through the $S^2$ at null infinity. The necessary condition to have 't Hooft-Polyakov monopoles is that the field configuration reach the vacuum at spatial infinity, so the case we have considered here is a particular case where the fields reach the vacuum at null infinity. 

Away from the charge insertions or without charge insertions at all, the potential $A_z(\infty,z,\bz)$ is holomorphic, as in the abelian case but with an important difference. Naively one could expect $A_{\bz}$ to be anti-holomorphic as in the $U(1)$ case, but since $F_{z\bz}=0$ then $\d_{z}A_{\bz}\neq0$.  It was noticed quite elegantly in \cite{He:2015zea} that in the limit where two gluons become soft, the result depends on the order of the limits, which in turns implies that the Kac-Moody algebra based on the group ${\cal G}$ associated to the $A_z$ is coupled to the one associated to the current $A_{\bz}$.   
This situation is similar to the one found in two-dimensional WZW models, where in order to guarantee a pair of independent holomorphic and an anti-holomorphic  Kac-Moody algebras it is needed to introduce a total derivative to the lagrangian, which supplements the equations of motions on the boundary at infinity. 
In four dimensions the natural candidate to be added to the Yang-Mills action (\ref{ggaction}) is a theta term,
\be
S_{\theta}=\frac{\theta}{32\pi^2}\int d^4x\tr F_{\mu\nu}{}^*F^{\mu\nu}
\ee
where ${}^*F^{\mu\nu}=\frac{1}{2}\epsilon^{\mu\nu\rho\sigma}F_{\rho\sigma}$ and this action in four dimensions is a total derivative that might be evaluated at $r=\infty$,
\be
\int d^4x\tr F_{\mu\nu}{}^*F^{\mu\nu}
=2\int_{\mathbb{R}\times S^2}dudzd\bz\,\tr\,\ep^{rmnl}(A_m\d_n A_l+A_mA_nA_l)\,.
\ee \footnote{In this expression, $r$ stands for the coordinate. }
This induces a Chern-Simons term on ${\cal I}^+$ while $\theta$ should be an integer multiple of $2\pi$ in order to preserve gauge invariance. At ${\cal I}^+$ the potential is a pure gauge $A_{z,(\bz)}=i\U\d_{z,(\bz)}\U^{-1}$ such that the theta action reduces to
\be
S_{\theta}=\frac{\theta}{16\pi^2}\int_{\mathbb{R}\times S^2}dudzd\bz\,\tr\,\ep^{rmnl}(\U\d_{m}\U^{-1}\,\U\d_{n}\U^{-1}\,\U\d_{l}\U^{-1})
\ee
The variation of the $\theta$ term is again a total derivative which lead us to 

\be
 \delta S_{\theta}=-\frac{\theta}{4\pi}\int d^2z\ep^{rvlm}(\U^{-1}\delta U\d_lA_{m})=-\frac{\theta}{4\pi}\int dvd^2z\ep^{rvlm}\d_v(\U^{-1}\delta U\d_lA_{m})\,.
\ee
Therefore we should sum up to the equations (\ref{veom}) the following term, 
\be
-\frac{\theta}{4\pi}\d_v(\ep^{rvlm}\d_lA_{m})
\ee
So the asymptotic equation (\ref{veom}) translate to 
\be
\d_{\bz} A_{z}\left(1+\frac{\theta}{4\pi}\right)+\d_z A_{\bz}\left(1-\frac{\theta}{4\pi}\right)+i g_{YM}[A_z,A_{\bz}]=g_{YM}\gamma_{z\bz}\int dv j^{E}_v\,,
\ee
or using the flat condition $F_{z\bz}=0$ and the current (\ref{pointcurrent}), we can rewrite,
\be
\d_{\bz} A_{z}\left(2+\frac{\theta}{4\pi}\right)-\d_z A_{\bz}\left(\frac{\theta}{4\pi}\right)=\frac{g_{YM}}{2}\sum_{k}\epsilon_k(z,\bz)\delta^2(z-z_k)\,,
\ee
At $\theta=0$, $A_z$ is holomorphic and at $\theta=-8\pi$ $A_{\bz}$ is anti-holomorphic away from the punctures on the sphere. Therefore, including a theta term to the Yang-Mills action, has the effect of decoupling the holomorphic and anti-holomorphic  Kac-Moody algebras at the boundary for particular values of the $\theta$ parameter, as we have mentioned at the introduction.

Due to the appearance of the holomorphic-antiholomorphic Kac-Moody algebras, the S-matrix elements would resemble correlations functions of some two-dimensional euclidean WZW model based on the gauge group ${\cal G}$ and therefore it would be interesting to look for a string theory realization of it in the way already suggested by Strominger\cite{Strominger:2013lka}.

The addition of a theta term breaks CP symmetry so it also would be interesting to see the consequences of this breaking on the S-matrix from the asymptotic symmetries perspective. Particularly, in the analysis of \cite{He:2015zea} it was important to consider CPT-invariant gauge transformations which allow to relate the asymptotic symmetries at past infinity with the ones at future infinity, so maybe it will be important to see how this matching conditions should be modified in the present case.
\\
\\
{\bf Acknowledgements}
I would like to thank to Freddy Cachazo and Humberto Gomez for carefully reading, correcting and making comments on this manuscript. I also thank to the Physics Department of the Universidad del Valle for the hospitality during the completion of this work.
\bibliography{EAB}{}
\bibliographystyle{utphys}

\end{document}